\documentstyle[prl,aps,twocolumn,epsf]{revtex}

\begin{document}
\draft
\title{Point contact tunneling in the fractional quantum Hall effect: 
an exact determination of the statistical fluctuations} 
\author{ H. Saleur${}^1$ and U. Weiss${}^2$}
\address{${}^1$ Department of Physics, University of Southern California,
Los Angeles, CA 90089-0484 }
\address{${}^2$ Institut f\"ur Theoretische Physik, Universit\"at Stuttgart,
D-70550 Stuttgart, Germany}
\date{Date: \today}
\maketitle
\begin{abstract}
In the weak backscattering limit, point contact tunneling between quantum Hall edges
is well described by a Poissonian process where Laughlin quasiparticles tunnel 
independently, leading to the unambiguous measurement of their fractional charges. 
In the strong backscattering limit, the tunneling is well described  by a Poissonian 
process again, but this time involving real electrons. In between, interactions 
create essential correlations, which we untangle exactly in this Letter. Our main 
result is an exact closed form expression for the probability distribution of the 
charge $N(t)$ that tunnels in the time interval $t$. Formally, this distribution 
corresponds to a sum of independent Poisson processes carrying charge 
$\nu e$, $2\nu e$, etc., or, after resummation, processes carrying charge 
$e$, $2e$, etc. In the course of the proof, we compare the
integrable  and Keldysh approaches, and find,  as a result of spectacular
cancellations between perturbative integrals, the expected agreement. 
\end{abstract}

\narrowtext

The study of dynamic fluctuations
is one of the most interesting aspects in the physics of mesoscopic
conductors. Most theoretical and experimental effort so 
far has concentrated on shot-noise \cite{BlanterButtiker}: 
shot noise is a consequence of the quantization of charge, and can be used to obtain 
information on a system, that would not be available through standard conductance
measurements. One of the most spectacular progress in this direction for instance has 
been the measurement of the charges of Laughlin quasiparticles in fractional
quantum Hall devices \cite{Saminayadar,DePicciotto,Reznikov}. In the ideal experiment
of interest here, one considers tunneling between two quantum Hall edges through a 
point contact. 
In the weak-backscattering limit, a picture where Laughlin quasiparticles tunnel 
independently becomes
exact to first order in the current, leading to a Poisson statistics for carriers of 
charge $\nu e$. In
the strong backscattering limit meanwhile, a similar picture holds but this 
time for electrons of charge $e$ \cite{KF,Wenetal,NewGlattli}. The complete noise 
in between these two regimes
has been calculated exactly in the field theory limit \cite{FLS,FS}, but this does not
shed too much light on the statistical fluctuations characterizing  the tunneling 
of the charges: questions of interest
include the nature of corrections to the Poissonian limit, and the striking 
cross-over between Laughlin quasiparticles and electrons. 

Study of higher moments of the current is a general way to get additional 
information on 
the physical processes involved in the shot noise: to give trivial examples, 
a vanishing of all higher cumulants 
would correspond to a Gaussian process, while finding that all the cumulants 
are equal would be indicative of a Poisson process. 

The most natural quantity to consider is the kth cumulant of the charge $N(t)$ 
that tunneled between the edges during the time $t$. In terms of the current, 
this cumulant can be expressed as
\begin{equation}
\left< N_j(t)\right>_c=\int_0^t dt_1\ldots dt_j \left<I(t_1)\ldots I(t_j)\right>_c \;.
\end{equation}
A knowledge of all the cumulants allows then (under reasonable regularity assumptions)
a reconstruction of the probability distribution of the variable $N(t)$ itself. 

So far, cumulants have been studied mostly in the 
case of  a pure partition noise, where noninteracting particles tunnel with a 
constant  probability $\tau$ \cite{oldpartition}. 
In this paper, we solve the much more difficult situation of the tunneling experiments 
in \cite{Saminayadar,DePicciotto,Reznikov}, where interactions are crucial. 

The calculation is done at temperature $T=0$, and in the field theory limit. 
The first condition 
corresponds essentially to the results reported in  
\cite{Saminayadar,DePicciotto,Reznikov}; it is not clear however that
the scaling regime was reached in these experiments, and more work in that direction 
is probably needed. Observe that, away from the scaling regime, results are nonuniversal, 
and strong\-ly dependent, for instance, on the shape of the point contact, 
hence limiting the interest of a theoretical analysis.

The solution of the tunneling problem in the scaling limit was first proposed in 
\cite{FLSletter}. It is based on the 
rather well known observation that the quantum field theory of interest - 
the boundary sine-Gordon (BSG) model - is integrable \cite{GZ}, but uses a new 
interpretation of this solution in terms of 
massless scattering \cite{ZZ} and a Boltzmann
equation for the integrable quasiparticles. The off-shoot is that current and 
noise can be calculated
using a model of massless particles with factorized scattering - an energy dependent 
version of the Haldane statistics \cite{Haldane}. The current for instance reads 
$I=\int_{-\infty}^\infty \!d\theta\,(\rho_+-\rho_-) \tau(\theta)$ where $\rho_\pm$ are 
densities of solitons and anti-solitons, $\theta$ the rapidity (parametrizing the 
energy by $E\propto e^\theta$),
$\tau(\theta)$ is a rapidity-dependent transmission coefficient, whose expression 
is given in \cite{FLS}, and we have set $e=h=v_F=1$. The
characteristic energy scale associated with the impurity is
$T_B\propto e^{\theta_B}$. While $\tau$ follows from the 
solution of the boundary Yang-Baxter equation \cite{GZ}, the densities are 
solution of a set of integral equations known as thermodynamic Bethe ansatz (TBA) 
equations.  The  noise has a  more complicated expression \cite{FS},
because thermal fluctuations of particle numbers at different energies are still 
coupled. However, in the $T=0$ limit, only shot noise remains. One finds 
therefore \cite{FLSletter}
\begin{equation}
<N_1>=tI=t\int_{-\infty}^A \!d\theta\,\rho(\theta)\tau(\theta) \;,  \label{current}
\end{equation}
together with \cite{FLS} 
\begin{equation}
<N_2>_c=t \int_{-\infty}^A \!d\theta\,
\rho(\theta)\left[\tau(\theta)-\tau^2(\theta)\right] \;.
\end{equation}
Here, only solitons contribute, because at $T=0$ the ground state is made of only 
one kind of particles, with
rapidities ranging from $-\infty$ (zero energy) to a voltage-dependent value of $A$
corresponding to the Fermi energy.

It was also discovered in \cite{FLS} that the noise is related with the current in a 
surprising way,
\begin{equation}
<I^2>=-{\nu\over 2(1-\nu)}{\partial I\over \partial\theta_B}\;, \label{flucdis}
\end{equation}
where what was called $<I^2>$ in \cite{FLS} corresponds to ${1\over t}<N_2>_c$ in the 
present notations. 
This relation is reminiscent of the fluctuation dissipation theorem (which of course 
does not hold in this non-equilibrium
situation). It occurs technically because the transmission probability $\tau$ behaves
essentially like a Fermi filling fraction, since it obeys
the basic identity
\begin{equation} \label{identity}
\tau(\theta)\left[\,1-\tau(\theta)\,\right]=-{\nu\over 2(1-\nu)} 
{\partial\over \partial\theta_B} \tau(\theta)\label{basic}  \;.
\end{equation}
In the following we introduce the new variable $
\tilde{\theta}_B\equiv -2 {1 - \nu \over \nu}\, \theta_B$.

We shall now use the logic of \cite{FLS} to compute all the other cumulants of $N(t)$ 
in the tunneling problem. For this, suppose first there were only one rapidity, 
and introduce for the corresponding quasiparticle a number $n=1$ if it
flips its charge when going through the impurity, $n=0$ if it does not. One has then, 
$\left< n\right>=\tau$. 
We can now consider the higher cumulants of the distribution of probability of $n$.
First, recall that
\begin{eqnarray}
<n_1>_c&=&<n>=\tau \;, \nonumber\\
<n_2>_c&=&<n^2>-<n>^2=\tau-\tau^2 \;, \nonumber\\
<n_3>_c&=&<n^3>-3<n><n^2>+ 2<n>^3\nonumber\\
&=&\tau-3\tau^2+ 2\tau^3\ldots  \; .
\end{eqnarray}
From the basic identity (\ref{identity}), we see that 
$<n_2>_c={\partial <n_1>_c\over \partial \tilde{\theta}_B}$. This result 
generalizes as follows
\begin{equation}
<n_j>_c={\partial^{j-1}\over (\partial \tilde{\theta}_B)^{j-1}}<n_1>_c\;.
\label{cumrel}
\end{equation}
The proof is quite simple. Consider the Fourier transform of the probability 
distribution of the variable $n$: $\hat{p} (k)=\int dn\, p(n) e^{ikn} $. 
Since $n$ takes only two values,
it reads simply $\hat{p}(k)=1+ \tau (e^{ik}-1)$. It follows that 
$\partial \hat{p}/\partial k=i\tau e^{ik}$ which we can rewrite as 
$\partial \hat{p}/\partial k= i \tau \hat{p}+i \tau(1-\tau)(e^{ik}-1)$. 
Use of the basic identity allows us to identify the latter term with a derivative 
with respect to $\theta'_B$. Thus we have
\begin{equation}
{\partial \hat{p}\over \partial ik}={\partial \hat{p}\over \partial \tilde{\theta}_B}
+\hat{p}<n_1>_c \;.
\end{equation}
The cumulants expansion $\ln \hat{p}=\sum_{j=0}^\infty <n_j>_c {(ik)^j\over j!}$ 
then leads us to Eq.~(\ref{cumrel}). 

Now, in the problem of interest, we have charges tunneling at arbitrary rapidities, 
so the quantities of interest are actually integrals. The average of  $N(t)$  
(the charge that went through the impurity in the time interval $t$) is related to 
the current calculated in \cite{FLS}. We have
\begin{equation}
<N(t)>=t \int_{-\infty}^A \!d\theta\,\rho(\theta)\tau(\theta)
= t I(\tilde{\theta}_B) \;,
\end{equation}
and more generally, 
\begin{equation}
<N_j(t)>_c= t\int_{-\infty}^A\; d\theta\, \rho(\theta) <n_j>_c(\theta) \;.
\end{equation}
If follows therefore that relation (\ref{cumrel}) also holds for the cumulants 
$<N_j>_c$. The Fourier transform of the probability distribution of the variable 
$N$ obeys then
\begin{eqnarray}\nonumber
{\partial\over \partial \tilde{\theta}_B}\ln \hat{P}(k) &=& t\sum_{j=1}^\infty 
{(ik)^j\over j!}{\partial^j\over (\partial \tilde{\theta}_B)^j} I(\tilde{\theta}_B) \\
&=&
t\left[ I(\tilde{\theta}_B+ik)-I(\tilde{\theta}_B)\right]\;.  \label{res1}
\end{eqnarray}
Determination of the integration constants leads then to 
\[
\ln \hat{P}(k)=2 {1-\nu\over \nu} t \int_{\theta_B}^\infty \!\!\!\!d\theta'_B \!
\left\{ I\left[\theta'_B-{ik\nu\over 2(1-\nu)}\right]
- I\left[\theta'_B\right]\right\}. 
\]
Using the expansions for the current derived in \cite{FLSbig} leads to the 
strong-backscattering expression for the Fourier transform of the probability 
\[
\hat{P}=\prod_{n=1}^\infty 
\exp\left\{ V t \left(e^{ikn}-1\right) \frac{a_n(1/\nu)}{n}
\left({V\over T'_B}\right)^{2n(1-\nu)/\nu} \right\} \;, 
\]
\begin{equation}
a_n(\nu) = (-1)^{n+1}
{\Gamma({3\over2})\Gamma(n\nu)\over \Gamma(n)\Gamma\left[{3\over 2}+n(\nu-1) 
\right]}\;.
\label{SBS}
\end{equation}
A similar expansion is possible in the weak-backscatter\-ing limit, giving rise to
\begin{eqnarray} 
\hat{P}&=&e^{\nu V ik  t} \\   \nonumber 
&\times&\prod_{n=1}^\infty 
\exp\left\{ \nu V t \left(e^{-ik\nu n}-1\right) {a_n(\nu)\over n}
\left({V\over T'_B}\right)^{2n(\nu-1)} \right\}  \;.
\end{eqnarray}
In these expressions, $T_B'$ is related to $T_B$ (and to the bare coupling constant) 
in a way made explicit in \cite{FLSbig}. Notice that the entire expansions
around the two limits are completely dual to each other.

The physical meaning of these expressions is quite fascinating. To elucidate it,
let us recall that for a  variable $X$ taking  integer values, this distribution 
is $p_X= m^X {e^{-m}\over X!}$. Here, $m$ is the average of $X$.
A simple property of the Poisson distribution is that all its cumulants are equal, 
so $m$ is the common value of all of them. Equivalently, 
$\hat{P}(k)=\exp[m\left(e^{ik}-1\right)]$. Suppose now that the charge observed 
in the time interval $t$ were the result of a Poisson process for particles of 
charge one going across the barrier, contributing a current $I_1$, 
plus a Poisson process for particles of charge two contributing 
a current $I_2$, etc. The final form of the Fourier transform of the probability 
distribution would then be
\begin{equation}\label{tbprob}
\hat{P}(k) = \prod_{n=1}^\infty \exp\left[  t \left(e^{ikn}-1\right)I_n/n\right] \;.
\end{equation}
This is extremely close to the expression (\ref{SBS}), the only difference being 
that the signs are not quite right:
while the classical process just considered requires all the $I_n$ to be positive, 
the coefficients in Eq.~(\ref{SBS}) are not. Certainly, the first contribution is 
indeed a Poisson process for the tunneling of electrons: but the would be tunneling 
of pairs of electrons (and of multiples thereof)
comes with the wrong sign, a result of quantum 
interference effects we will analyze in more details below. 

The weak-backscattering expansion is of the form
\begin{equation}\label{wbs}
\hat{P}(k) = e^{\nu Vikt}\prod_{n=1}^\infty \exp\left[  t  
\left(e^{-ik\nu n}-1\right)\tilde{I}_n/n\right] \;.
\end{equation}
The meaning is quite similar: apart from the overall factor which corresponds to the 
current in the absence of tunneling (and therefore, absence of fluctuations
at $T=0$), observe that the exponents are now of the form $e^{-ik\nu n}$. 
The factor $\nu$ occurs because now we are dealing with tunneling of  Laughlin  
quasiparticles and multiples thereof; the minus sign,
because their tunneling diminishes the current, instead of building it up
as in the strong backscattering limit. The situation for the signs is now dependent 
upon $\nu$. Using the inversion identity for gamma functions, the sign of 
$\tilde{I}_n$ for $n>1$ is the one of $\cos n\pi\nu$, and therefore
positive for the first few values of $n$ provided $\nu$ is small enough. 
The image of clusters of Laughlin quasiparticles tunneling independently with a 
classical Poisson process is therefore quite good for bundles with modest $n$, 
in the small $\nu$ limit. As $\nu$ goes to zero, 
all the coefficients are positive, but the fluctuations disappear, as one reaches 
the classical limit:
$$
\hat{P}=\exp\left\{ \nu V ik  t
\left[1-\sum_{n=1}^\infty {\Gamma(n-1/2)\over 2\sqrt{\pi}\, n!} 
\left({V\over T'_B}\right)^{-2n}\right]\right\}  \; .
$$

Notice also the remarkable fact that each of the ``means'' of the would be 
Poissonian distributions are, apart from a common factor, pure power laws of the 
dimensionless coupling constant $(V/T'_B)^{2(\nu-1)}$. 
This result precludes the $\tilde{I}_n$ from being all positive: in this case indeed, 
the expansions would have a singularity on the real axis, which is nonphysical,
except for $\nu=0$.

We stress here that the Poissonian aspect occurs only after integration over the 
rapidity variables, thanks to the basic identity (\ref{basic}). At any given rapidity 
(that is, energy), the noise is simply a partition noise, and 
$$
\hat{p}= 1 +\tau\left(e^{ik}-1\right)
$$
is {\sl not} Poissonian, except for very small values of $\tau$. 
Remarkably therefore, the interactions have actually made 
the final result simpler than in the noninteracting case!

The above results rely entirely on the integrable approach. We now
would like to reconsider them in the light of the 
nonequilibrium Keldysh or Feynman-Vernon method. 
To do so, it is convenient to use the correspondence 
of the BSG model with the Schmid model \cite{Schmid} of dissipative 
quantum mechanics.
This model describes a quantum Brownian particle coupled to an Ohmic heat bath and 
moving in a tilted cosine potential. The strong-backscattering limit corresponds to
the tight-binding (TB) limit of this model.
In the TB representation, the dimensionless viscosity $K$ is related to the
filling fraction $\nu$ by $K=1/\nu$,
and the bias energy $\epsilon/2\pi$ corresponds to the voltage $V$.
The scaled point contact interaction $T_B'$ is related to  the bare transfer 
matrix element $\Delta$ and the cutoff $\omega_c$ as $T^{\prime\,2-2/\nu}_B =
[\,\pi^{2K}/\Gamma^2(K)\,]\,\Delta^2/\omega_c^{2K}$ \cite{Weiss}.

To proceed, suppose we had 
a general Poissonian TB transport model with transition rates $\gamma_m^{\pm}$ describing
direct forward $(+)$ and backward $(-)$ transitions by $m$ states.
Assuming statistically independent transitions, 
the Fourier transform of the
probability distribution $\hat{P}(k)=\sum_n e^{ikn}\,P(n)$ at time $t$ is found as
\begin{equation}\label{bpmt}
\hat{P}(k) = \prod_{n=1}^\infty \exp[t (e^{ikn}-1)\gamma_n^+
+ t(e^{-ikn} -1)\gamma_n^-] \;.
\end{equation}

At $T=0$, backward moves are absent, and we have
\begin{equation}\label{bpm}
\hat{P}(k) = \prod_{n=1}^\infty \exp[t (e^{ikn}-1)\gamma_n^+] \;.
\end{equation}
The expressions (\ref{tbprob}) and (\ref{bpm}) are equal, but the 
interpretation for the terms $n>1$ is somewhat different. While (\ref{tbprob}) 
describes joint transport
of two particles, three particles, etc., through a single point contact, the expression
(\ref{bpm}) represents nonsequential (coherent) tunneling of a single particle 
across one intermediate TB state, two intermediate TB states, etc. We
shall now show, using the Keldysh or Feynman-Vernon formalism, that
this Poissionian description of the TB model at long times is in fact exact.
Moreover, it turns out to be possible to calculate the rates at $T=0$
perturbatively, confirming the results of the integrable approach at 
the lowest orders, and sheding light on their general structure. 

It is indeed straightforward to
derive the exact series expression for the generating functional $\hat{P}(k,t)$
of the above dissipative model \cite{Weiss}.
The Laplace transform $\hat{P}(k,\lambda)$ is the analog of an isobaric ensemble of
a classical gas of charges stringed along the forward and backward path. 
The charge interaction factor for $m$ time-ordered charges $\{u_j=\pm 1\}$ on the 
forward path $q$ and $l$ time-ordered charges $\{v_i=\pm 1\}$ on the backward path 
$q'$ is
\begin{eqnarray*}
{\cal F}[q_m,q'_l] &=& \exp\bigg\{
\sum_{j=2}^m\sum_{i=1}^{j-1} u_j u_iQ(t_j -t_i) \\[-1mm]
 && \!\!\!\!\!\!\!\!\!\!\!\!\!\!\!\!\!\!\!\!\!\!\!\!
+ \sum_{j=2}^l \sum_{i=1}^{j-1} 
v_j v_iQ^\ast(t'_j- t'_i)
-\sum_{i=1}^l \sum_{j=1}^m v_i u_j Q(t'_i - t_j) \bigg\} ,
\end{eqnarray*}
where $Q(t)= 2K\ln[(\omega_c/2\pi^2 T)\sinh2\pi^2 T t]  +i\,\pi K\,{\rm sgn}\,t$.
For a charge sequence contributing to $P(n)$, we have the constraints
$\sum_j u_j = 2n$ and $\sum_i v_i =2n$.

The expression (\ref{bpmt}) follows from a cluster decomposition
of the series for $\hat{P}(k,\lambda\to 0)$. The clusters are the
$\lambda$-independent (irreducible) path sections which start and end in 
diagonal states of the reduced density matrix. By definition, the clusters are
noninteracting and therefore separated by a factor $1/\lambda$. 
Path segments with intermediate visits of diagonal states are reducible, i.e., 
factorize into clusters of lower order.
Observe that after subtraction of the reducible components always an irreducible
part is left.
The ``transition rates'' $\gamma_n^\pm$, which  do not necessarily have to be 
positive, can be identified as the sum of all clusters which interpolate
between the (arbitrary) diagonal state $m$ and the diagonal state $m\pm n$.
All the charge sequences contributing to the rate can be split up in subsets $\alpha$,
$\gamma_n^\pm = \sum_\alpha\gamma_n^{(\alpha)\pm}$. 
In each subset, the arrangement $\{u_i,v_j\}$ is fixed and 
time-ordered along the Keldysh contour.   
The property $Q(t-i/2\pi T) =Q^*(t)$ ensures detailed balance for the  
dynamics, actually  already for every subset $\alpha$, 
$\gamma_n^{(\alpha)-} = e^{-\epsilon n/2\pi T}\,\gamma_n^{(\alpha)+}$ \cite{Weiss}.

At $T=0$, the clusters formally appear as the series 
\begin{equation}\label{expan1}
\gamma_n^+ = (\epsilon/2\pi)\,x^{2n}\sum_{l=0}^\infty x^{2l}f_n^{(l)} \;,
\qquad x = {\Delta\over\epsilon}\left({\epsilon\over\omega_c}\right)^K \;,
\end{equation}
in which the coefficients $f_n^{(l)}$ are given in terms of a sum of 
$2(n+l)-1$--fold integrals,
each of them representing a particular arrangement of $2(n+l)$ charges.
Using detailed balance for the subsets,
$\gamma_n^{(\alpha)-}=0$,   
a multitude of relations between the various integrals of the same order
can be derived \cite{SW}.
We have analyzed in detail all clusters up to  order $x^6$.
It turns out that  $\gamma_1^+$ is of order $x^2$,
and there are no contributions of order  $x^4$ and $x^6$. Similarly,
$\gamma_2^+$ is found to be of order $x^4$, and there is no contribution of order
$x^6$. This indicates that there are formidable cancellations between the various
charge sequences, and only those contribute where all charges $\{u_j\}$ and
$\{v_i\}$ are positive. From this, it is easy to  conjecture, that in the series 
for $\gamma_n^+$, at $T=0$, only the $m=0$ term is left,
\begin{equation}\label{expan2}
 \gamma_n^+ = (\epsilon/2\pi)\,x^{2n} f_n^{(0)}, 
\end{equation}
in agreement with the result from the integrable approach.
Namely, a moment relation analogous to (\ref{cumrel}),
$< N_j(t)>_c =  (x\,\partial/\partial x)^{j-1}_{}<N_1(t)>_c $,
 would follow directly from (\ref{bpm}) with 
(\ref{expan2}).  
A general proof of (\ref{expan2}) for arbitrary $n$ is still missing:
it may well require an analysis of the problem similar to the one in 
Refs.~\cite{BLZ}, \cite{fs95}.

As a final remark, we observe that (\ref{expan2}) leads to the 
relation $f_n^{(0)} = [2^{1-K}\pi/\Gamma(K)]^{2n}_{}\, a_n(K)/n$, 
and thus allows us to construct the full probability distribution for the displacement 
of the quantum Brownian particle too \cite{SW}.

\end{document}